\documentclass[aps,prd,twocolumn,superscriptaddress,preprintnumbers, longbibliography]{revtex4-1}

\usepackage{graphicx}
\usepackage{bm}
\usepackage{times}
\usepackage{slashed}
\usepackage{amssymb}
\usepackage{color}
\usepackage{slashed}
\usepackage{lipsum}

\usepackage{multirow}
\usepackage{amsmath}
\usepackage{array}
\usepackage{varwidth}
\usepackage{float}
\usepackage{xcolor}
\usepackage[%
  colorlinks=true,
  urlcolor=blue,
  linkcolor=blue,
  citecolor=blue
]{hyperref}

\usepackage{latexsym}
\begin{document}

\title{Modeling in-ice radio propagation with parabolic equation methods}

\author{S.~Prohira}%
 \email{prohira.1@osu.edu}
 \affiliation{Department of Physics, Center for Cosmology and AstroParticle Physics (CCAPP), The Ohio State University, Columbus OH, USA}

 \author{C.~Sbrocco}%
 \email{sbrocco.6@buckeyemail.osu.edu}
 \affiliation{Department of Physics, Center for Cosmology and AstroParticle Physics (CCAPP), The Ohio State University, Columbus OH, USA}

\author{P.~Allison}
 \affiliation{Department of Physics, Center for Cosmology and AstroParticle Physics (CCAPP), The Ohio State University, Columbus OH, USA}
\author{J.~Beatty}
 \affiliation{Department of Physics, Center for Cosmology and AstroParticle Physics (CCAPP), The Ohio State University, Columbus OH, USA}
 \author{D.~Besson}
 \affiliation{University of Kansas, Lawrence, KS, USA}
 \affiliation{National Research Nuclear University, Moscow Engineering Physics Institute, Moscow, Russia}
 \author{A.~Connolly}
  \affiliation{Department of Physics, Center for Cosmology and AstroParticle Physics (CCAPP), The Ohio State University, Columbus OH, USA}
 \author{P.~Dasgupta}
  \affiliation{Universit\'{e} Libre de Bruxelles, Brussels, Belgium}
  \author{C.~Deaconu}
\affiliation{Enrico Fermi Institute, Kavli Institute for Cosmological Physics, Department of Physics, University of Chicago, Chicago, IL, USA}

  \author{K.D.~de~Vries}

\affiliation{Vrije  Universiteit  Brussel,  Brussel,  Belgium}

  \author{S.~De~Kockere}
  \affiliation{Vrije  Universiteit  Brussel,  Brussel,  Belgium}
  \author{D.~Frikken}
  \affiliation{Department of Physics, Center for Cosmology and AstroParticle Physics (CCAPP), The Ohio State University, Columbus OH, USA}
  \author{C.~Hast}
 \affiliation{SLAC National Accelerator Laboratory, Menlo Park, CA, USA}
 \author{E.~Huesca~Santiago}
 \affiliation{Vrije  Universiteit  Brussel,  Brussel,  Belgium}
 \author{C.-Y.~Kuo}
 \affiliation{National Taiwan University, Taipei, Taiwan}
 \author{U.A.~Latif}
 \affiliation{University of Kansas, Lawrence, KS, USA}
   \affiliation{Vrije  Universiteit  Brussel,  Brussel,  Belgium}
  \author{V.~Lukic}
  \affiliation{Vrije  Universiteit  Brussel,  Brussel,  Belgium}
 \author{T.~Meures}
 \affiliation{University of Wisconsin-Madison, Madison, WI, USA}
 \author{K.~Mulrey}
 \affiliation{Vrije  Universiteit  Brussel,  Brussel,  Belgium}
 \author{J.~Nam}
 \affiliation{National Taiwan University, Taipei, Taiwan}

\author{A.~Nozdrina}
\affiliation{University of Kansas, Lawrence, KS, USA}
\author{J.P.~Ralston}
\affiliation{University of Kansas, Lawrence, KS, USA}
 \author{R.S.~Stanley}
 \affiliation{Vrije  Universiteit  Brussel,  Brussel,  Belgium}
\author{J.~Torres}
 \affiliation{Department of Physics, Center for Cosmology and AstroParticle Physics (CCAPP), The Ohio State University, Columbus OH, USA}
 \author{S.~Toscano}
  \affiliation{Universit\'{e} Libre de Bruxelles, Brussels, Belgium}
  \author{D.~Van~den~Broeck}
  \affiliation{Vrije  Universiteit  Brussel,  Brussel,  Belgium}
  \author{N.~van~Eijndhoven}
  \affiliation{Vrije  Universiteit  Brussel,  Brussel,  Belgium}
 \author{S.~Wissel}
 \affiliation{Departments of Physics and Astronomy \& Astrophysics, Institute for Gravitation and the Cosmos,   Pennsylvania State University, University Park, PA, USA}
\affiliation{California Polytechnic State University, San Luis Obispo CA, USA}
\collaboration{Radar Echo Telescope}

\begin{abstract}
We investigate the use of parabolic equation (PE) methods for solving radio-wave propagation in polar ice. PE methods provide an approximate solution to Maxwell's equations, in contrast to full-field solutions such as finite-difference-time-domain (FDTD) methods, yet provide a more complete model of propagation than simple geometric ray-tracing (RT) methods that are the current state of the art for simulating in-ice radio detection of neutrino-induced cascades. PE are more computationally efficient than FDTD methods, and more flexible than RT methods, allowing for the inclusion of diffractive effects, and modeling of propagation in regions that cannot be modeled with geometric methods. We present a new PE approximation suited to the in-ice case. We conclude that current ray-tracing methods may be too simplistic in their treatment of ice properties, and their continued use could overestimate experimental sensitivity for in-ice neutrino detection experiments. We discuss the implications for current in-ice Askaryan-type detectors and for the upcoming Radar Echo Telescope; two families of experiments for which these results are most relevant. We suggest that PE methods be investigated further for in-ice radio applications.
\end{abstract}
 
\maketitle

\section{Introduction} Accurate modeling of radio wave propagation in the ice is essential for experiments seeking to detect in-ice neutrino interactions at the highest energies. These experiments (past, present, and planned) consist of radio frequency (RF) antennas above or buried in the ice from 0-200\,m below the surface, that seek to detect radio waves a) emitted by the cascade produced by an ultra high-energy neutrino interaction in the ice (the Askaryan effect\cite{askaryan_orig,rice, anita,ara, arianna, rno}) or b) reflected from the ionization deposit left in the wake of the cascade (the radar echo method~\cite{krijnkaelthomas, radioscatter,krijn_radar_18, t576_run2}). Due to the very low flux of neutrinos with energies in excess of $10^{15}$\,eV (PeV), large volumes must be instrumented in order to detect a statistically significant number of neutrinos. Radio-based methods for in-ice neutrino detection exploit the relative transparency of ice at radio frequencies, which allows for radio detectors to instrument very large volumes with sparse apparatus. In many cases, a radio signal will be propagating at least in part through a region with a changing index of refraction ($n(z)$ for the depth coordinate $z$), typically the top 100-200\,m of an ice sheet, called the firn.

\begin{figure}[H]
  \centering
  \includegraphics[width=0.45\textwidth]{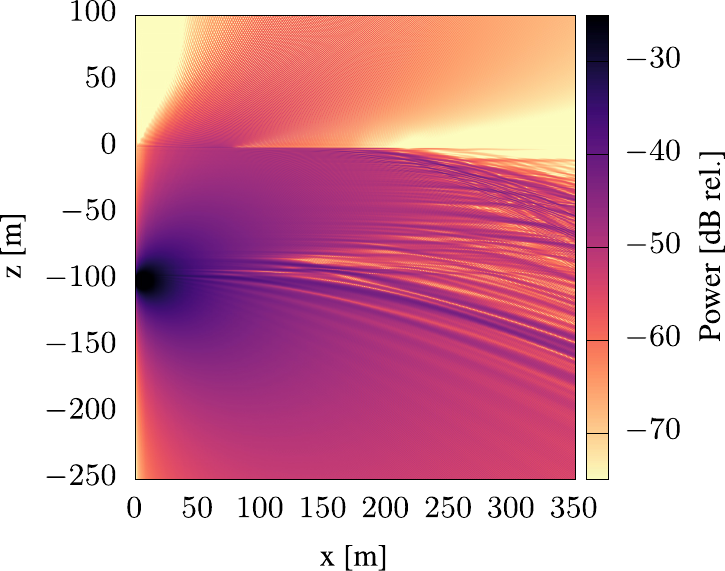}
  \caption{An example of radio propagation modeled by the parabolic solver, using the index of refraction profile from the south pole, derived from SPICE~\cite{spice} core data in the top 100 meters and a functional fit below. For this figure we model 350\,MHz continuous wave radio from a dipole source 100\,m below the surface.}
  \label{ex_map}

   \end{figure}

Typically, such propagation is treated using a formalism of ray tracing (RT)~\cite{rice_index_of_refraction,araConstraintFirst, nuradiomc,nuradioreco, raytrace_github_uzair,arasim_github}, where a ray is propagated along a path with discrete straight-line segments according to the rules of geometric optics, applicable when the wavelength is much smaller than any feature size, and wave effects such as diffraction and interference can be neglected. The direction of each segment is dictated by the given refractive index profile, which is typically a functional fit to data. These methods are computationally efficient, reaching solutions on the order of ms. This allows for use in Monte Carlo simulations. Recent studies~\cite{horizontal_propagation_fdtd} have shown, however, that more complete modeling of the firn, simulated using finite-difference-time-domain (FDTD) methods, can replicate observed signal features in data~\cite{horizontal_propagation_barwick, horizontal_propagation_exp_2018} that ray-tracing solutions can not. Moreover, it has been shown that density fluctuations in the ice can complicate propagation. A density profile ($\rho(z)$) can be converted to an index of refraction profile ($n(z)$) via the expression $n(z) = 1 + 0.845\rho(z)$~\cite{densityToN}. These fluctuations can result in unexpected amplitudes as a function of source and receiver geometry (if the exact density profile is not known, which is typical), as well as horizontally propagating modes for certain source and receiver configurations. These effects are most pronounced at shallow depths, near the surface of the ice, where density variations are maximized. So, while RT methods have many strengths, they do not provide a full picture of radio propagation, and therefore result in limitations on event reconstruction, both for neutrino arrival direction and primary energy, for which the properties of the received signal spectrum are of paramount importance.

 The FDTD formalism is robust~\cite{fdtd}, but is computationally expensive. The entire purpose of using radio to instrument large volumes of ice in search of ultra high energy neutrino interactions is to cover a massive volume with minimal apparatus, and detect signals across great distances, yet it is intractable (and in some cases impossible) to simulate wide band time-domain signals over kilometer scale baselines using FDTD methods. Therefore in this article we explore whether simple parabolic equation (PE) solvers, similar to those used for decades in atmospheric propagation studies~\cite{levy} and undersea acoustic studies~\cite{tappert1977parabolic,fleck1976time,feit1978light, brock1977modifying,thomson1983wide}, can be applied  to the problem of in-ice radio wave propagation.~\footnote{Though EM and acoustic waves differ (transverse vs. longitudinal), the form of the wave equation is the same in both cases. In the former, the field quantity is the electric or magnetic field, and in the latter, the pressure.}  An example of a PE solution for 350\,MHz continuous-wave radio from a transmitter 100\,m beneath the ice is shown in Figure~\ref{ex_map}. We find that, while prone to phase errors due to the rate of change of the $n(z)$ profile, PE methods generally provide a more accurate modeling of the spectral content of a signal (using FDTD as a baseline) than RT methods. A better modeling of the spectral content of simulated signals may improve energy and arrival direction reconstruction relative to methods currently being used. 

 The article is organized as follows. We first introduce the PE method, and show validation studies against an open source FDTD package (\texttt{meep}~\cite{meep}) over a small domain where FDTD routines are tractable.  We compare also to ray tracing solutions in the range of their validity. We then extend the study to simulate time-domain signals at long baselines with PE and RT methods. We conclude by discussing next steps and the implications for current and future experiments.

 \section{In-ice simulation methods}
 There are 3 simulation methods that we discuss in this paper: the parabolic equation (PE), finite-difference time-domain (FDTD), and ray-tracing (RT). In this section, we introduce these, focusing on the PE methods which are being introduced to the problem of in-ice radio detection of neutrinos for the first time.
 \subsection{Ray tracing methods}
 In-ice ray tracing is currently the standard simulation technique for experiments seeking to detect in-ice neutrino interactions using radio. RT methods take the infinite frequency limit, and solve for the multiple paths that a signal can travel from source to receiver, often designated ``direct'' (the signal that travels from a source to receiver on an arc without intercepting the surface) and  ``reflected'' (the signal that reflects from air/ice interface). 
 In general, a particular transmitter--receiver geometry will admit one or two ray path solutions. In all cases, rays travel along curved paths when traversing the firn. RT methods have analytic solutions for some forms of the index of refraction profile and in general are numerical solvers that can provide vertexing (identification of the location of the RF source) based on time-difference-of-arrival (TDOA) of different antennas. If a receiver is in a geometry where both direct and reflected signals are present, a TDOA between direct and reflected can be an additional powerful variable for vertex resolution.
 RT methods are computationally efficient, and some implementations can use arbitrary (i.e. non-functional) $n(z)$ profiles. In all cases, however, they rely on geometric optics to calculate propagation. In this article, we use RT for a functional $n(z)$ profile, as well as a data-driven $n(z)$ profile as explained below. To optimize the RT simulation, the step size scale factor was reduced until convergence, with time delay precision better than 0.1 ns.

 \subsection{Finite-difference time-domain methods}

 These methods solve Maxwell's equations numerically on a spatial grid in the time domain. They are a standard for time-domain electromagnetic modeling in antenna design, interference analysis, and numerous other applications. Because they involve approximations only in the discretization of the problem, they are accurate as long as the grid spacing is sufficiently small for the frequencies of interest (a good approximate rule is a minimum of 10 cells per wavelength). In this work we have simulated everything on a 5\,cm grid for both FDTD and the PE methods described below, which is sufficient for frequencies up to 600\,MHz, in excess of what we simulate in the work. The primary draw-backs to FDTD methods are the large memory requirements (which for cylindrical volume of $R$ by $Z$ with resolution $r$ scales with $RZr^2$) and the long computational time (which scales with $RZr^3$).

 \subsection{Parabolic equation methods} The parabolic equation (PE) is an approximation of the full wave equation which can be solved to allow for stepping solutions for field propagation. Simply, this means that in order to calculate the electric field at some distance from your source (the range coordinate, here denoted $x$) you only need the electric field at the previous range step. This stepping, as opposed to solving for the electric field across the entire simulation domain at each time step (FDTD) results in significantly lower computational cost. It does, however, come at the expense of accuracy. The parabolic equation is only valid within a certain angular range of the propagation direction (called the paraxial direction), and only (at least for the simple simulation shown here) includes forward propagating fields. 

Our implementation of the PE method (a wide-angle, split-step solver, described below) is based on the aforementioned standard reference works for in-air EM and in-water acoustic PE solvers. We extend these standard methods with a new, modified split-step approximation motivated by the in-ice problem. We present the derivation and details of the PE in Appendix~\ref{pe}, including the updated split-step approximation. Wide-angle means that the solution is valid at wider angles to the paraxial direction (up to ~90$^{\circ}$ depending on frequency). Split-step means that at each step in range, the solution is split into diffractive and refractive components, solved for sequentially (this is equivalent, in this case, to splitting the solution at each step into time and frequency domain components). The cylindrically-symmetric field $\psi(x, \theta, z)$, polarized along $\theta$, is solved for via Equation~\ref{eq:parabolic_final}, which shows the reduced field $u(x, z)=\sqrt{x}e^{-ik_0x}\psi$ (assuming an $e^{i\omega_0t}$ time dependence) for a range step $x+\Delta x$. Use of the reduced field in this form allows for the solution of the wave equation for $\psi$ (see Equation~\ref{eq:u3d}) in a convenient form. It is only dependent on the previous range step, is cylindrically symmetric, and valid in the far field.

 Denoting by $\mathcal{F}$ and $\mathcal{F}^{-1}$ a forward and backward Fourier transform, respectively, and $k_0$, $k_z$ the reference wavenumber and the Fourier space wavenumber, respectively, the field at range step $x+\Delta x$ is given by

 \begin{multline}
  \label{eq:parabolic_final}
  u(x+\Delta x, z)=\textrm{exp}\bigg[ik_0\left(n\sqrt{1+ \frac{1}{n_0^2}}-\sqrt{1+\frac{n^2}{n_0^2}}\right)\Delta x\bigg]\\
  \mathcal{F}^{-1}\bigg\{\textrm{exp}\bigg[-ik_0\Delta x\sqrt{1-\frac{k_z^2}{k_0^2}}+1\bigg]\mathcal{F}\big\{u(x, z)\big\}\bigg\},     
\end{multline}
where $n=n(x, z)$ and $n_0$ is a reference index of refraction corresponding to the reference wavenumber.  This is built into a FOSS python code~\cite{paraprop_github} available on GitHub.

\begin{figure}[t]
  \centering
  \includegraphics[width=0.44\textwidth]{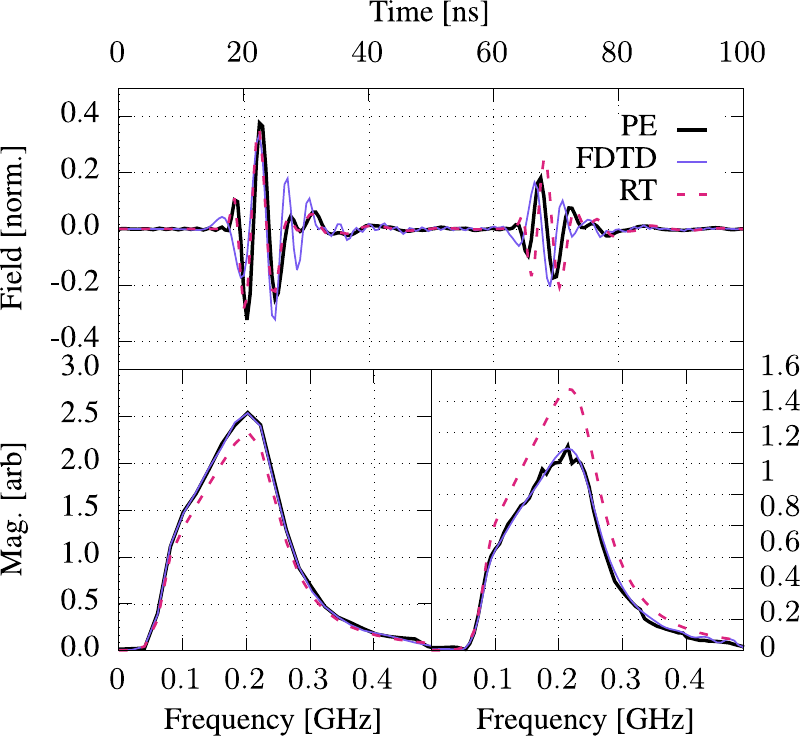}
  \caption{Time domain (top) and frequency domain (bottom) comparisons between FDTD (thin solid line), PE (thick solid line), and RT (dashed line) for a source at $x, z = (0, -30)$\,m and a receiver at $x, z = (100, -25)$\,m, with a functional $n(z)$ profile corresponding to a fit to South Pole data. The two frequency domain plots correspond to the direct (left) and reflected (right) pulses in the time domain signal.}
     \label{td_compare_100_25_smooth}

 \end{figure}

   The PE method as described is a spatial solver. To implement the PE solver in the time domain, we decompose the spectrum of a time-domain source pulse into the individual Fourier modes and simulate each mode using the associated complex amplitude.  We then synthesize a received spectrum in the same way at a particular receiver position, and take the inverse transform to arrive at a received signal. An example of this is shown in Figure~\ref{td_compare_100_25_smooth}, where a time domain impulse, band-limited from 90-250\,MHz using a 4th order Butterworth filter, is simulated in FDTD, PE, and RT from a transmitter at 30\,m depth. The received signal for a receiver 100\,m away in range at a depth of 25\,m, using a functional form for $n(z)$ at the South Pole ($n(z)=A-B e^{-Cz}$, with $A=1.78$, $B=0.43$, and $C=0.0132$\,m$^{-1}$), is shown. Clearly visible in all 3 time-domain waveforms resulting from the 3 propagation methods are the direct and reflected signals, as the signal will be seen both directly, and reflected from the ice/air boundary for this geometry. We also show the spectrum for the direct and reflected pulses separately. We will discuss the properties of this waveform in detail in a later section, where we make more comparisons between the 3 methods for a variety of transmitter/receiver locations, and for different $n(x,z)$ profiles.

 The two main differences between the parabolic equation and the FDTD solutions are the source definitions and boundary conditions. In a typical FDTD simulation, a physical source is defined, an excitation applied to this source, and the fields are calculated as a function of position and time. For the PE methods, a source is defined as the full reduced function (see Appendix~\ref{source}) along a single range step (e.g. definition of a source at range step 0 means defining the reduced function at every point along $z$ at that range step). Boundary conditions in FDTD simulations are handled by solving the solutions numerically for varying material properties on a grid. Boundary conditions, including boundary roughness, are handled in the PE in various ways. For the present studies, we implement a flat surface with air above and an $n(z)$ profile below corresponding to either a uniform density, or the density profile of the south pole, but we implement this ``boundary'' as part of a complete $n(z)$ profile. That is, boundary conditions are not put in ``by-hand'' as they sometimes are in PE applications, but instead, the fields are reflected from the surface as though it were a density fluctuation within a continuous $n(z)$ profile. This results in good agreement with FDTD for direct and reflected amplitudes when using the split-step approximation shown here. Further study is needed to verify the validity of our implementation of this ``boundary'' treatment, but comparison to FDTD implies that it is a reasonable approach.

 \section{Results}

 In this section we show several comparisons between the various methods, for different geometries and $n(x, z)$ profiles. The first section presents the results for a selection of receiver geometries and a functional profile for the ice at the South Pole, which is the same profile at every range step $x$, hence $n=n(z)$. The functional form of $n(z)$ is a 3 parameter fit to the measured data as used in radio codes. The second section shows the results for the same geometries, albeit with data-derived $n$ profiles. We show the case where we apply the same $n(z)$ profile to every range step for 2 different density data samples taken at the South Pole Ice Core Experiment (SPICE) core~\cite{spice}. These profiles are obtained from experimental data, but applying them uniformly in range may result in waveguide-like behavior, overestimating the effect of the firn, so some care is needed in the interpretation of the results. To address this, we also show the results for the case where we linearly interpolate between the two SPICE cores along the range domain of the simulation (hence $n=n(x,z)$), a scenario justified by experimental data, discussed below. In the comparison figures in this article, we compare normalized amplitudes to emphasize relative timings and spectral content. All signals are aligned via cross-correlation in the time domain.

 The source for all waveforms is a vertically-polarized (axis along $z$ at $x=0$) dipole. This is implemented in FDTD as a dipole current source, and in PE as an initial source model (detailed in Appendix~\ref{source}).  To get a time-domain waveform out of RT~\cite{araPermittivity,ariannaRecoSnow}, one needs 4 parameters: the time of arrival of the first pulse, the time of arrival of the second pulse, the Fresnel coefficient for the reflection from the surface (if there is a reflected path), and the ``launch angle'' of the transmitted signal(s). We use this launch angle to select the appropriate output pulse(s) at 1\,m from the FDTD source.
 We then delay the first and second pulses by their associated time delays, and scale the second pulse by the (complex) Fresnel coefficient and by the relative path length difference between the first and second pulse to obtain an amplitude (and appropriate phase shift for pulses undergoing total internal reflection). Ice signal absorption is not currently included in any of the methods shown here, but will be investigated in future work. 
 In general, we find that PE consistently outperform RT (qualitatively) for agreement with FDTD in the time domain, and most noticeably in the frequency domain.

 \subsection{Functional $n(z)$ profile}

 An example of a received signal has been shown in Figure~\ref{td_compare_100_25_smooth}, as described above. In this configuration, the source and receiver are at roughly the same depth, and the vertical receiver is 100\,m horizontally displaced from the source. In such a configuration, both the direct and reflected pulses are evident.
 The path from source to receiver is not described by a straight line in either case. The wave front, instead of radiating away with uniform velocity, as would be the case in a uniform medium, progresses with a depth-dependent phase velocity, distorting its shape. For a smoothly-varying index of refraction that increases with depth, this results in lower velocities deeper down, higher velocities near the surface. Thus a horizontally-propagating wavefront will be `bent' down from horizontal when moving through such a depth-dependent index of refraction profile. Small scale fluctuations will cause local variations in the phase velocity, further distorting the wave front.
 
\begin{figure}[h]
  \centering
    \includegraphics[width=0.44\textwidth]{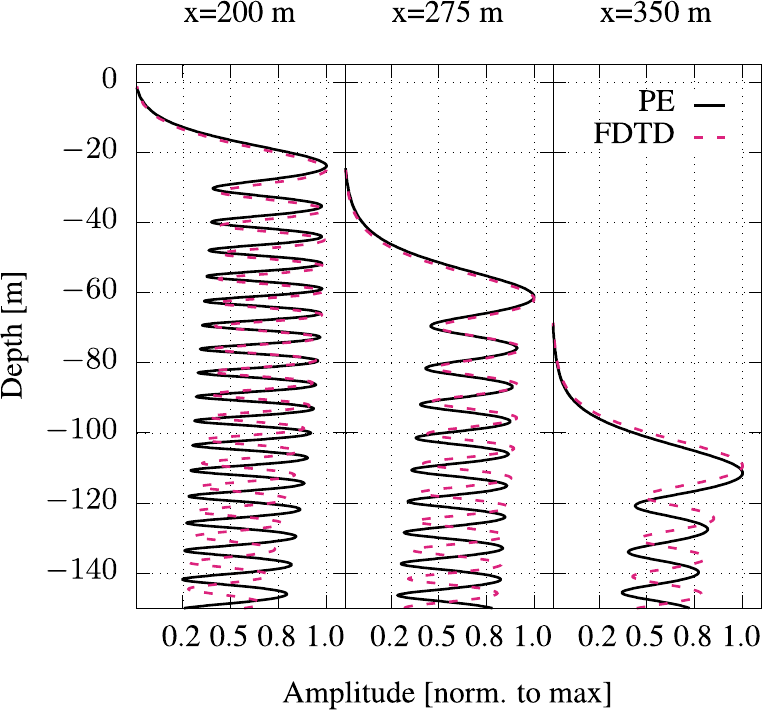}
   \caption{A comparison for discrete range steps for the maximum value of the received field, for PE (solid line) and FDTD (dashed line). This is for a continuous-wave, 135\,MHz signal, with the direct and reflected interference pattern evident. From left to right, range steps of 200\,m, 275\,m, and 350\,m are shown. The curves have been normalized to the peak $x$-axis value to assist comparison.}
   \label{fig:zProfile}
 \end{figure}

In Figure~\ref{fig:zProfile} we show the peak field at discrete range steps in $x$ for PE and FDTD for a 135\,MHz continuous-wave signal. We note the following: First, the envelope of the two shows good agreement. The depth of the troughs in the interference pattern is a measure of the relative strength of the direct and reflected signals as they interfere, and here we see that the $z$ profiles, normalized to their peak amplitude, show good similarity throughout the depth. If, for example, the surface reflection coefficient were higher, the peak/trough ratio in the interference pattern would be increased, since the reflected signal strength would be closer to the direct. In addition to the depth of the troughs, the general shape of the envelopes (that is, the overall trend with $z$ of the left and right extent of the interference pattern) is similar, indicating that the source function in PE is a decent approximation of the dipole simulated by the FDTD methods. The envelopes would differ if, for example, one were a directive antenna (which would see higher amplitudes at the peak of the antenna's gain pattern). Second, the location of the ``shadow zone'' boundary, that is, where the fields `turn on' in depth, is well matched between the methods at all ranges. This is important for studies about horizontal propagation, and will be discussed in more detail below. Third, we see that there is some disagreement in the phase of the interference pattern between the methods as a function of depth. We found that some of this disagreement can be mitigated by the use of an appropriate reference wavenumber (discussed later), but some disagreement persists. Some methods~\cite{brock1977modifying} have introduced modified indices of refraction to solve this issue for the acoustic case, but as yet, we have not found a suitable method to do so for the large values of $n(z)$ in ice. We believe this is a solvable problem. In any case, this interference pattern does not seem to measurably affect the time-domain signals being analyzed in this paper; perhaps a detailed phase analysis would shed some light, but for now this is beyond the scope of our introductory study.

\begin{figure}[H]
  \centering
    \includegraphics[width=0.44\textwidth]{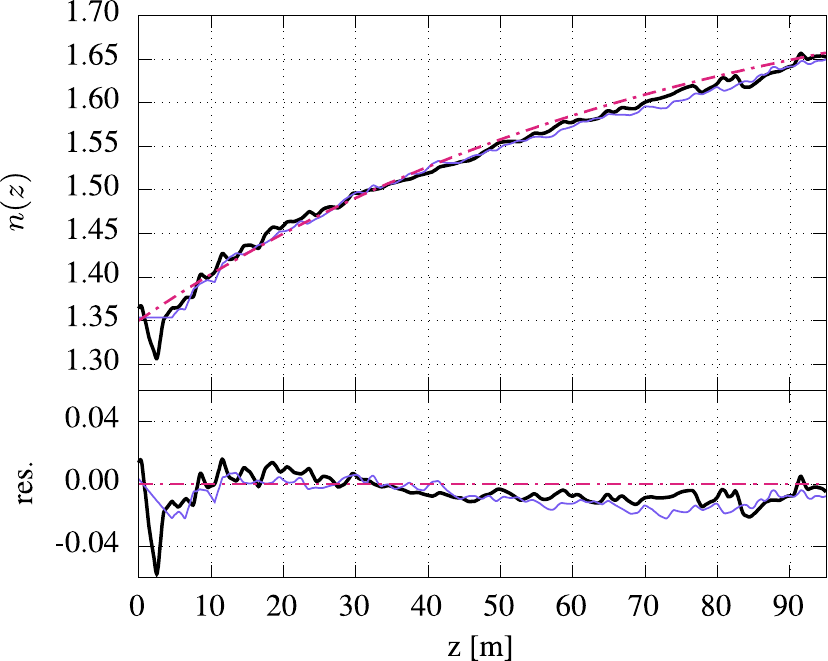}
   \caption{The $n(z)$ profile (top panel) as measured by the two SPICE core~\cite{spice} firn holes (core 1, thick solid line, core 2 thin solid line), along with the functional fit (dashed line) to these data used for ray tracing codes at the south pole. Residuals are shown below.}
   \label{fig:nzprofile}
 \end{figure}
 
 \subsection{Measured $n(x, z)$ Profile}

 The South Pole Ice Core Experiment (SPICE)~\cite{spice} extracted a 1751-meter long ice core at South Pole. While drilling the main core, two shallow samples were taken from the top $\sim$100\,m of the ice sheet (aka 'firn'). These shallow cores were measured every half meter for density, and from that, an index of refraction can be calculated. Figure~\ref{fig:nzprofile} shows the data for the 2 cores along with the functional fit used in RT. It is clear that there are differences between the cores and the functional fit, but also between the two cores themselves. The dataset does not provide measurement errors on these density measurements, though density measurements from other sites in Antarctica and Greenland~\cite{hawley2008rapid,waisDensity} show similar variations with depth. We use these measurements as-is for purposes of investigating how changes in the density profile affect propagation, and whether this can be modeled with PE methods.

 RT methods do not (in general) take into account reflections from small-scale density fluctuations, which seem to contribute to the overall change in the propagated signal. We find, as one might expect, that when density fluctuations are taken into account in PE, the agreement between FDTD and PE improves, while the agreement between RT and the wave methods gets worse. We simulate 3 different scenarios for the data-driven case:

 \begin{enumerate}
 \item SPICE core \#1 at all $x$
 \item SPICE core \#2 at all $x$
 \item $n(x, z)$ with each slice in $x$ linearly interpolated between SPICE core \#1 at $x=0$, and core \#2 at $x=300$\,m
 \end{enumerate}

 \begin{figure}[H]
   \centering

  \includegraphics[width=0.44\textwidth]{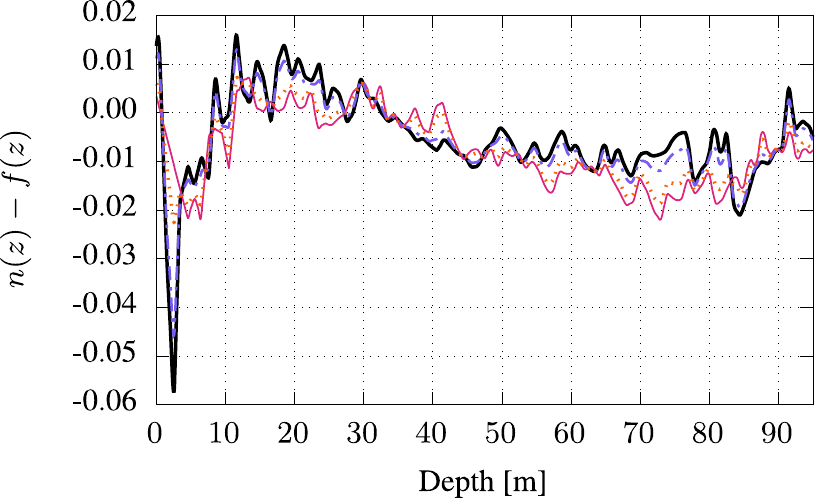}
   \caption{The residuals of the $n(z)$ profile and the functional fit $f(z)$, as measured by the two SPICE core firn holes (core 1, thick solid line, core 2 thin solid line), along with profiles linearly interpolated at $1/4$ (dashed) and $3/4$ (dotted) interval between these two, used in the interpolated $n(x, z)$ case.}
   \label{fig:nzprofile2}
 \end{figure}

Note that in the following comparisons, the third case (linear interpolation) is shown compared to the RT result from SPICE core \#2. This is due to limitations in the RT implementation used, for which $n$ cannot vary with range. This limitation is present in all known RT methods for in-ice radio propagation at the time of this writing.

 From ground-penetrating radar (GPR) data~\cite{spikes2004variability,arcone2005phase} we see that fluctuations in $n(z)$ remain more-or-less constant over hundreds of meters to kilometers. Therefore, for this present study, we have decided that 300\,m is a reasonable range over which to linearly interpolate between two different profiles in order to simulate a more realistic ice profile. In Figure~\ref{fig:nzprofile2} we show the 2 cores and 2 simulated cores at $1/4$ and $3/4$ of the way in range, used to make the full $n(x,z)$ map. The full $n(x,z)$ is likely the truest to the actual configuration of the firn, though truly local defects are not included, and the 300\,m correlation length is a data-driven approximation. Future models can perhaps use the density information from GPR surveys to back out a more accurate ice density model. In reality, all three of these methods may over-or underestimate the amplitude or the discretization of the $n(x,z)$ profile in either range or depth, and the present study is simply meant to show that changes in the $n(z)$ profile can  alter the received signal spectrum with respect to idealized ice models. We note that this $n(x,z)$ model also, in all three cases, preserves the deviation in the top $\sim$20\,m of the ice that has been observed experimentally~\cite{horizontal_propagation_barwick} to cause time-of-arrival inversions for horizontally propagating signals.

\begin{figure}[h]
   \centering
     \includegraphics[width=0.44\textwidth]{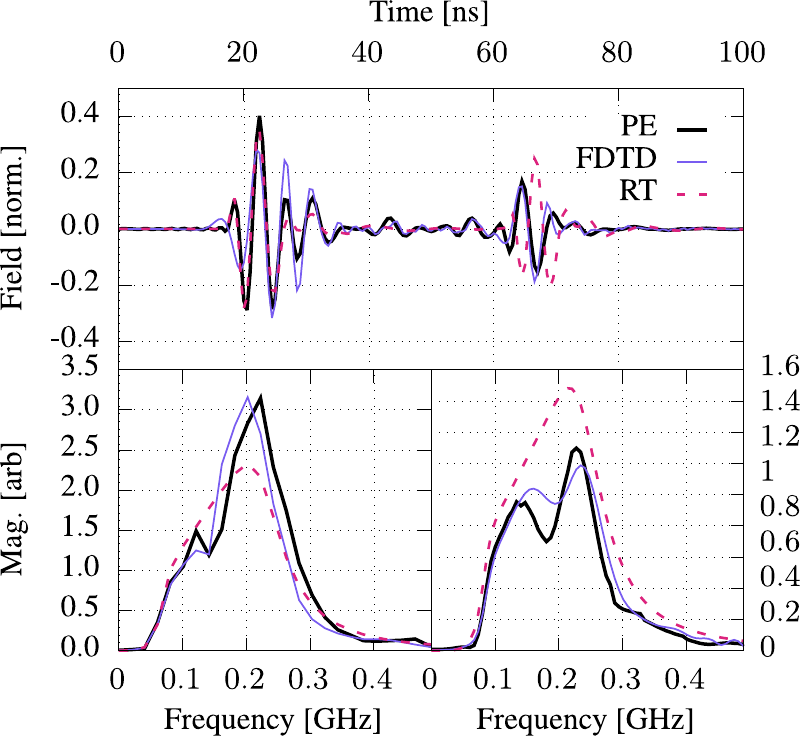}
\caption{Time domain (top) and frequency domain (bottom) comparisons between FDTD (thin solid line), PE (thick solid line), and RT (dashed line) for a source at $x, z = (0, -30)$\,m and a receiver at $x, z = (100, -25)$\,m, with an $n(z)$ profile corresponding to the first of 2 SPICE core firn samples. The two frequency domain plots correspond to the direct and reflected pulses in the time domain signal.}

     \label{td_compare_100_25_core1}

 \end{figure}

\begin{figure}[h]
   \centering
     \includegraphics[width=0.44\textwidth]{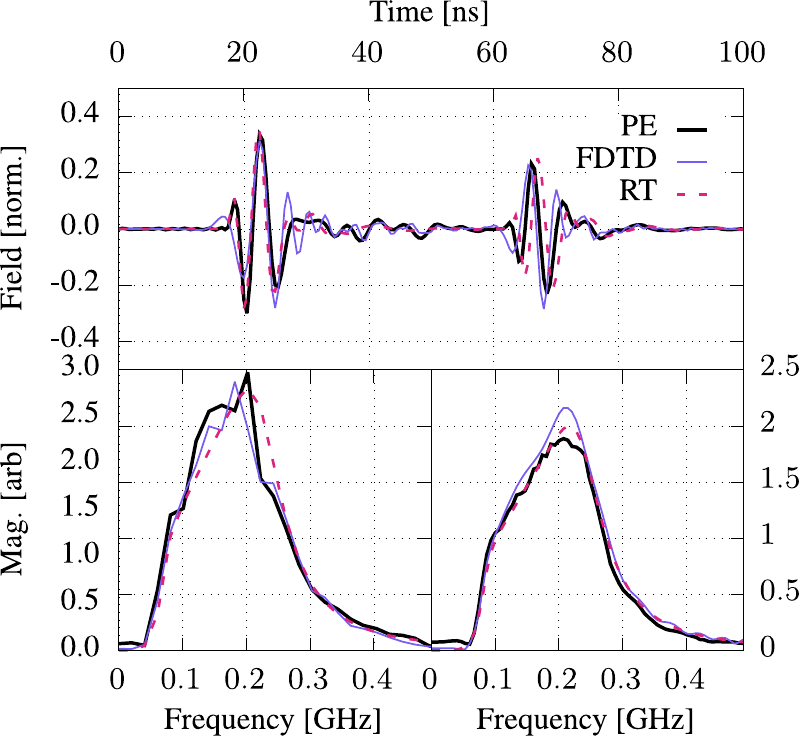}
\caption{Time domain (top) and frequency domain (bottom) comparisons between FDTD (thin solid line), PE (thick solid line), and RT (dashed line) for a source at $x, z = (0, -30)$\,m and a receiver at $x, z = (100, -25)$\,m, with an $n(z)$ profile corresponding to the second of 2 SPICE core firn samples. The two frequency domain plots correspond to the direct and reflected pulses in the time domain signal.}

     \label{td_compare_100_25_core2}

 \end{figure}

\begin{figure}[h]
   \centering
     \includegraphics[width=0.4\textwidth]{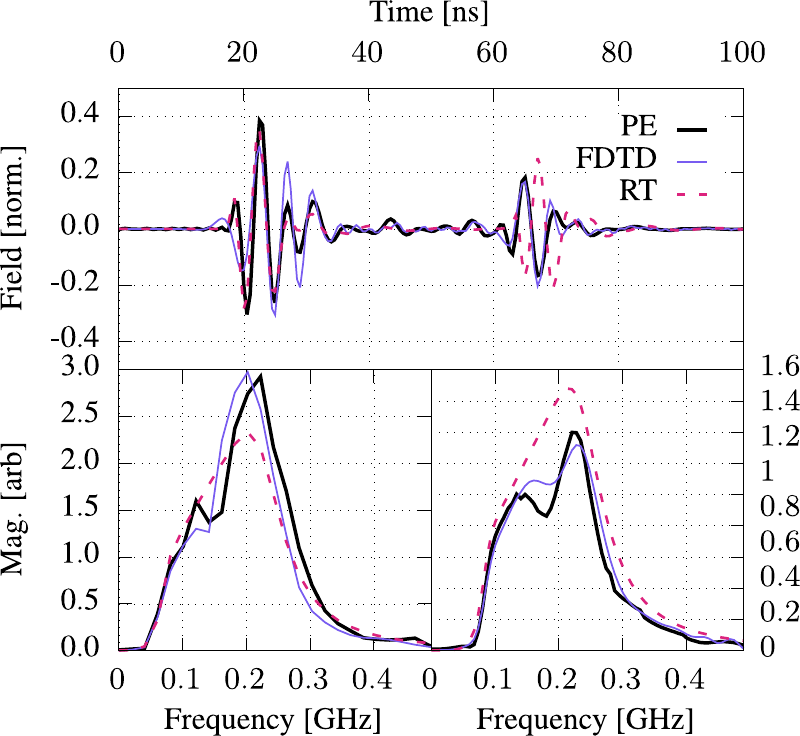}
\caption{Time domain (top) and frequency domain (bottom) comparisons between FDTD (thin solid line), PE (thick solid line), and RT (dashed line) for a source at $x, z = (0, -30)$\,m and a receiver at $x, z = (100, -25)$\,m, with an $n(x,z)$ profile corresponding to a linear interpolation between the first SPICE core at $x=0$ and the second SPICE core at $x=300$\,m.}
     \label{td_compare_100_25_avg}

 \end{figure}

 First we show the same configuration as Figure~\ref{td_compare_100_25_smooth} (source=(0, -30)\,m, RX=(100, -25)\,m), for each of the 3 configurations listed above in Figures~\ref{td_compare_100_25_core1}, \ref{td_compare_100_25_core2}, and \ref{td_compare_100_25_avg} respectively. Notice the agreement in waveform shape, timing of the direct and reflected pulses, and the relative amplitudes of the two pulses in each.  For this configuration, where the first and second pulses are clearly distinguishable, we show individual spectra for the direct and reflected pulses below the time domain waveform, with each spectrum corresponding to the pulse above it. The agreement between all three methods for the functional profile is quite good; discrepancies become apparent between RT and the wave methods as $n(x,z)$ deviates from the pure functional form.  We do not explore in this article whether modifications can be made to existing RT methods to capture the physics of the wave methods.

\begin{figure}[h]
      \centering
        \includegraphics[width=0.44\textwidth]{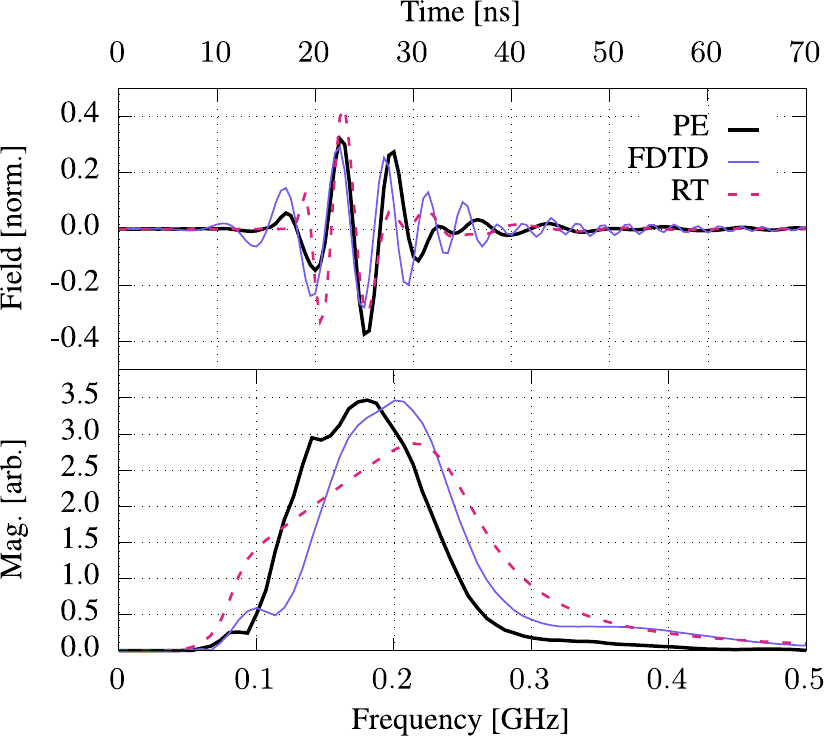}
\caption{Time domain (top) and frequency domain (bottom) comparisons between FDTD (thin solid line), PE (thick solid line), and RT (dashed line) for a source at $x, z = (0, -100)$\,m and a receiver at $x, z = (250, -2)$\,m, with an $n(z)$ profile corresponding to the first of 2 SPICE core firn samples.}
     \label{td_compare_250_2_core1}

 \end{figure}

\begin{figure}[h] 
   \centering
     \includegraphics[width=0.44\textwidth]{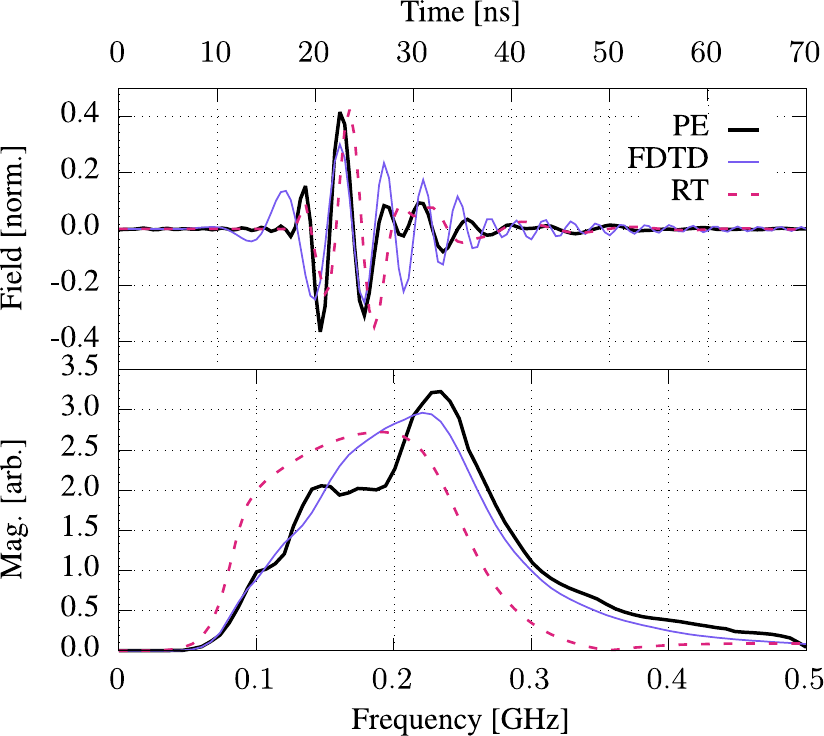}
  \caption{Time domain (top) and frequency domain (bottom) comparisons between FDTD (thin solid line), PE (thick solid line), and RT (dashed line) for a source at $x, z = (0, -100)$\,m and a receiver at $x, z = (250, -2)$\,m, with an $n(z)$ profile corresponding to the second of 2 SPICE core firn samples.}
     \label{td_compare_250_2_core2}

 \end{figure}

\begin{figure}[h] 
   \centering
     \includegraphics[width=0.44\textwidth]{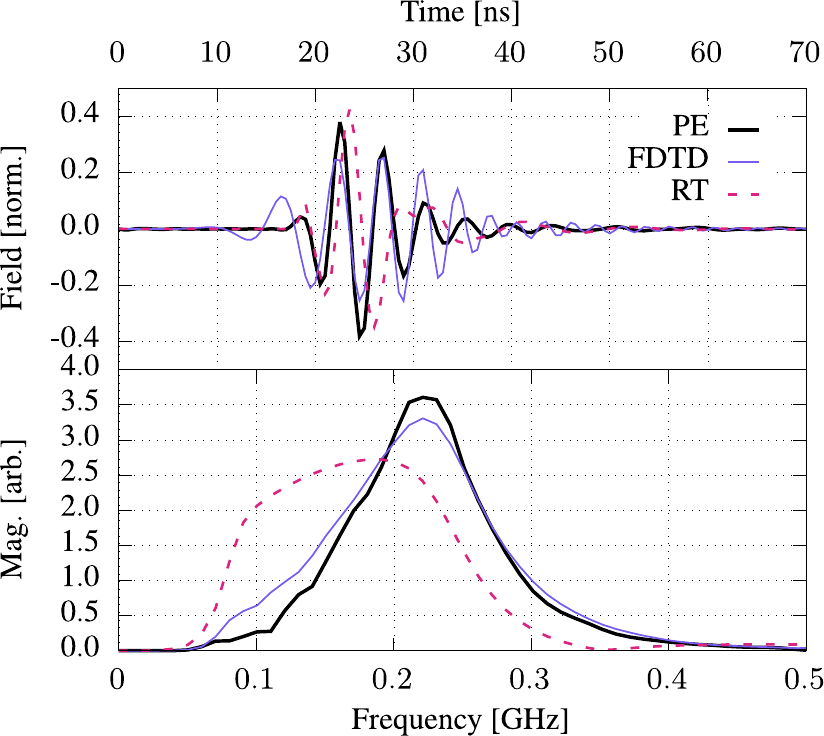}
  \caption{Time domain (top) and frequency domain (bottom) comparisons between FDTD (thin solid line), PE (thick solid line), and RT (dashed line) for a source at $x, z = (0, -100)$\,m and a receiver at $x, z = (250, -2)$\,m, with an $n(x,z)$ profile corresponding to a linear interpolation between the first SPICE core at $x=0$ and the second SPICE core at $x=300$\,m.}
     \label{td_compare_250_2_avg}

 \end{figure}

   We then show plots for the 3 different $n$ configurations as above, but this time for a source at (0, -100)\,m and RX at (250, -2)\,m, in Figures~\ref{td_compare_250_2_core1}, \ref{td_compare_250_2_core2}, and \ref{td_compare_250_2_avg} respectively. We see that the difference between RT and the wave methods becomes pronounced, particularly in the frequency domain. There is some discrepancy between PE and FDTD in the frequency domain as well, but the overall shapes are qualitatively similar. The wave solutions lose more low frequency power as they advance spatially and temporally, and in some cases become more peaked, which can be understood intuitively: the $n(x, z)$ profile acts as a rough lens. As the field propagates through this lens, a) low frequency power is lost more rapidly simply due to aperture, and b) certain frequencies are (de)focused more than others for particular geometries. The envelope of the RT spectrum is always bounded by the envelope of the initial pulse (by construction, since the eventual signal is constructed from the initial pulse, shifted in time, scaled, and in some cases phase shifted), but the wave methods allow for physical, frequency dependent effects to be more readily modeled. We also note that the alignment of the traces, performed with cross-correlation, becomes poorer at this distance, as the shapes of the pulses diverge. In these cases, it is instructive to look at the frequency domain plots to see the spectral content, as the time-domain signals can show deceptive (dis)agreement. This is most evident in the pulses in the interpolated $n(x,z)$ case (Figure~\ref{td_compare_250_2_avg}), where the time-domain signals look similar ``by-eye'', but the spectrum shows clear peaking, and is quite similar for FDTD and PE. The loss of low frequency power is significant; at 120\,MHz for example, the discrepancy between PE and RT is $\sim$15\,dB in power. 

   \subsection{Shadow-zone propagation}

\begin{figure}[h]
      \centering
        \includegraphics[width=0.44\textwidth]{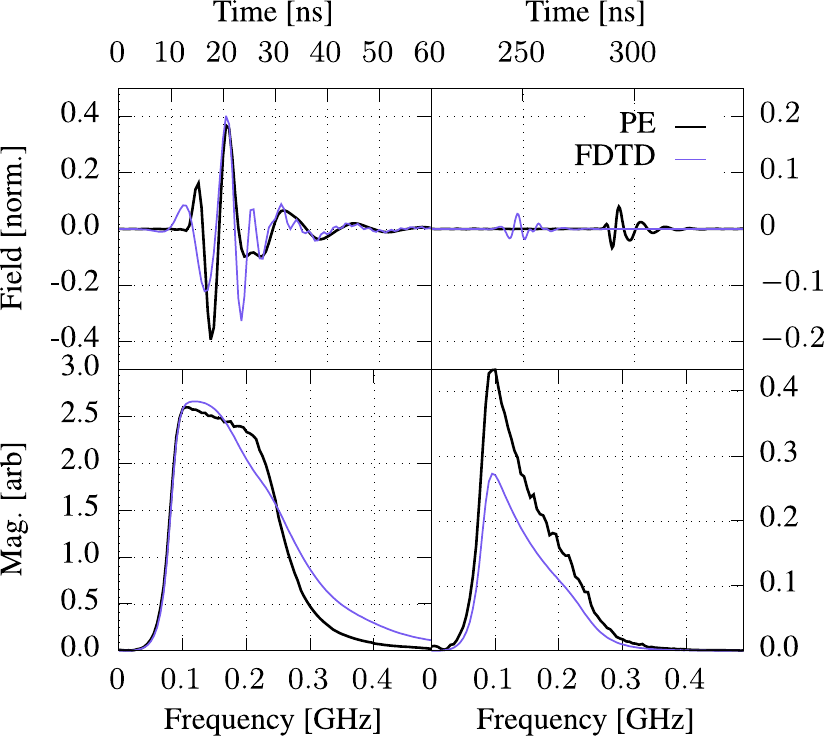}
     \caption{Time domain (top) and frequency domain (bottom) comparisons between PE (thick solid line), and FDTD (thin solid line) for a source at $x, z = (0, -30)$\,m and a receiver at $x, z = (250, -2)$\,m, with a functional $n(z)$ profile for the South Pole. The columns in the time and frequency domain plots correspond to the first and second signals, respectively.} 
     \label{fig:shadow1}

 \end{figure}

\begin{figure}[H]
       \centering
         \includegraphics[width=0.44\textwidth]{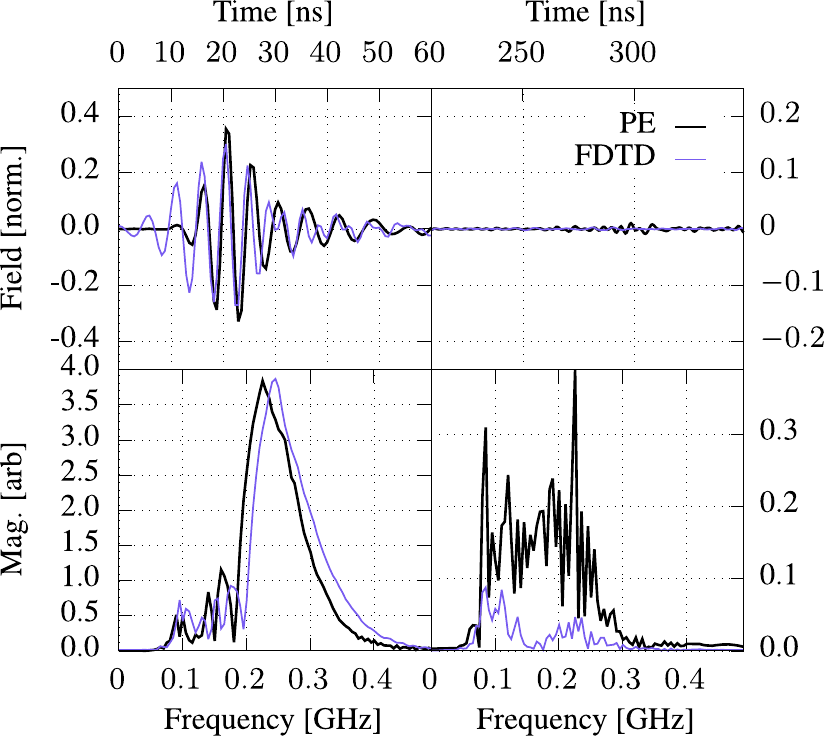}
     \caption{Time domain (top) and frequency domain (bottom) comparisons between PE (thick solid line), and FDTD (thin solid line) for a source at $x, z = (0, -30)$\,m and a receiver at $x, z = (250, -2)$\,m, with $n(x,z)$ corresponding to a linear interpolation between SPICE core 1 at $x=0$ and SPICE core 2 at $x=300$. The columns in the time and frequency domain plots correspond to the first and second signals, respectively.} 
     \label{fig:shadow2}

 \end{figure}

   A distinct advantage of wave methods is that they can be used to simulate propagation in regions where RT methods find no solution, the so-called ``shadow'' or ``forbidden'' zone. This is the region beyond the most distant caustic of the field that bends from source to receiver (in Figure~\ref{ex_map} a receiver at $(x,z)=325,-5$ would find itself in the shadow zone). Wave or field methods  solve for the field at each point in the domain, so there are no forbidden zones. We show in Figures~\ref{fig:shadow1} and \ref{fig:shadow2} a comparison between FDTD and PE for a receiver position for which ray-tracing does not permit a solution, because it is in the shadow zone. Figure~\ref{fig:shadow1} is for a functional $n(z)$ profile, and Figure~\ref{fig:shadow2} is for the linearly interpolated $n(x,z)$ profile. There is decent agreement between FDTD and PE in time and frequency in both cases (albeit with discrepancy in arrival time of the second pulse for the functional $n(z)$ parameterization, see Figure~\ref{fig:shadow1}, top right panel). The source of this discrepancy is not known, but investigation of such signals could identify the specific processes (see e.g.~\cite{shadowZoneTutorial} for a tutorial on boundary phenomena) behind these shadow zone signals. Since shadow zone signals have been observed~\cite{horizontal_propagation_barwick, horizontal_propagation_exp_2018}, studied with FDTD~\cite{horizontal_propagation_fdtd}, and investigated phenomenologically~\cite{lahmannReco}, any new simulation must include such effects, which the PE method does (at least qualitatively). Further investigation into these effects is underway.

 \subsection{Larger simulation domain}

\begin{figure}[h]
   \centering
     \includegraphics[width=0.44\textwidth]{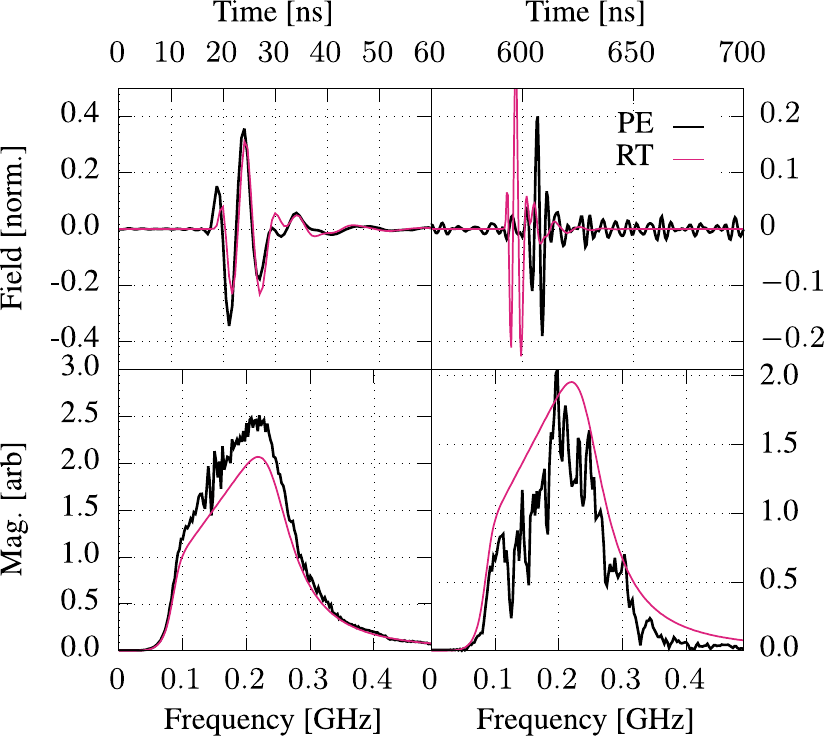}
     \caption{Time domain (top) and frequency domain (bottom) comparisons between PE (thick solid line), and RT (thin solid line) for a source at $x, z = (0, -1050)$\,m and a receiver at $x, z = (1350, -120)$\,m, with a $n(z)$ profile corresponding to the second of 2 SPICE core firn samples. The columns in the time and frequency domain plots correspond to the direct and reflected signals, respectively.} 
     \label{td_compare_1350_120}

 \end{figure}

 The results presented so far have been shown for a domain tractable for FDTD simulations, however, true neutrino vertices are expected to be much deeper in the ice. The PE solver can produce time domain waveforms over arbitrarily large domains on manageable timescales. This code computes a time domain waveform for the bandwidth in this paper on a 1.5\,km x 1.5\,km domain in about 0.5 CPU thread hours (compared to $\sim$40 for FDTD), and has not been optimized. The Fourier synthesis time-domain method is well-suited for parallelization, and is being explored. We note that, while these times offer improvement over FDTD, they are still much longer than RT solutions. Therefore PE solvers would not (at this stage) replace RT for all operations, but only for those in which greater spectral accuracy is sought. In this section we present some waveforms comparing RT and PE on a larger domain, and summary figures showing how the spectrum of a waveform changes as a function of viewing angle for a receiver, with implications for experimental design and neutrino event reconstruction.
 We place a transmitter at 1050\,m below the ice surface and model the same band-limited pulse from a dipole source as before. This time, however, the domain is large for FDTD, so we only compare PE and RT. We use a SPICE profile for $n(z)$. In Figure~\ref{td_compare_1350_120}, the receiver is placed at a horizontal displacement of 1350\,m, and a depth of 120\,m beneath the ice surface. The four panels correspond to the direct (left) and reflected (right) pulses in the time (top) and frequency (bottom) domains. There are several things specifically to note. First, the direct pulse has a spectrum that agrees fairly well between PE and RT, with some distortion. Second, the reflected pulse is far more distorted than the direct pulse, as this has traversed more of the firn en route to the receiver. For this particular geometry, the spectrum largely traces out the envelope of the frequency-domain spectrum of the RT signal, but this may not be true in general. Third, there is a time offset between PE and RT for the reflected signal. Further work is needed to investigate the source of this offset.

  \begin{figure}[h]
    \centering
      \includegraphics[width=0.44\textwidth]{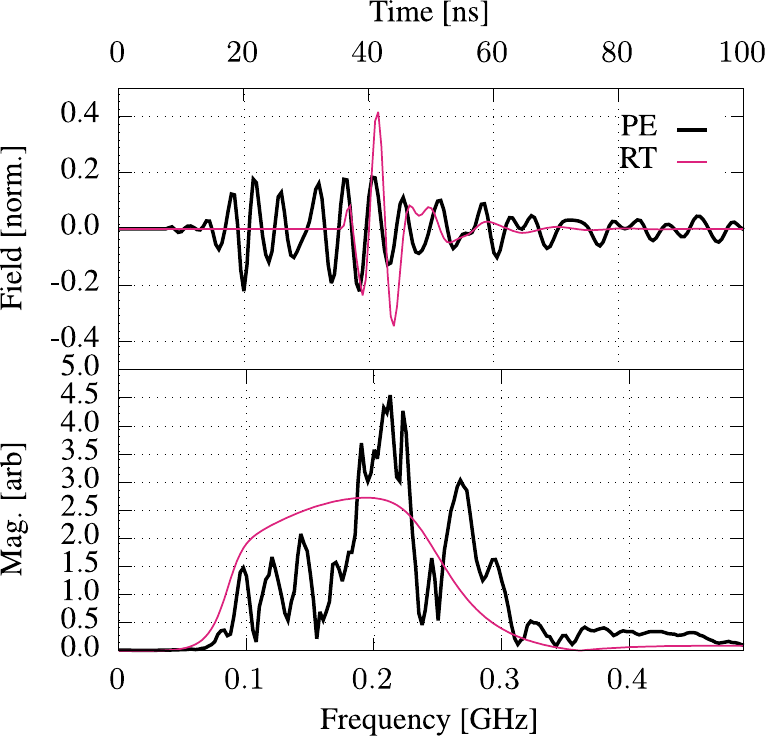}
     \caption{Time domain (top) and frequency domain (bottom) comparisons between PE (thick solid line), and RT (thin solid line) for a source at $x, z = (0, -1050)$\,m and a receiver at $x, z = (1350, -2)$\,m, with a $n(z)$ profile corresponding to the second of 2 SPICE core firn samples. }
     \label{td_compare_1350_2}

 \end{figure}
 
 Next, Figure~\ref{td_compare_1350_2} plots the same setup but for a receiver just 2\,m beneath the ice surface. Note the disagreement between RT and PE. The time-domain signals are aligned by their maximum cross-correlation value, but the pulses are so dissimilar that this alignment is largely meaningless here. There is evidence of interference from the surface, but also the signal in PE is more strongly peaked, indicating that there could be evidence of waveguide-like phenomena at this particular distance. The envelopes of the spectra also do not agree well. 

\begin{figure}[ht]
  \centering
    \includegraphics[width=0.44\textwidth]{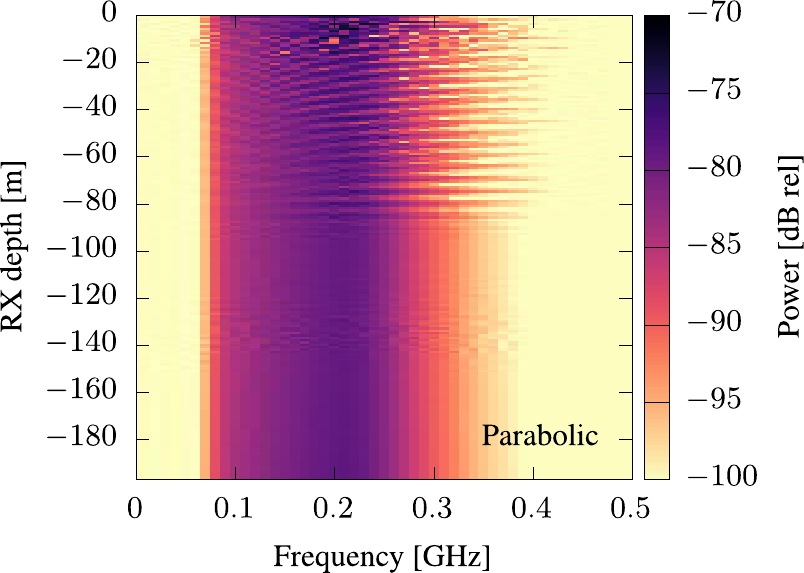}
     \caption{Spectrum of the first 100ns of a received signal for varying receiver (RX) depth using parabolic equations. This is for a transmitter buried 1050\,m in the ice, displaced 1\,km in range from the receivers. The scale is dB (power) relative to the transmitter output.}
     \label{fig:spectrumDepth}

   \end{figure}

 Finally, we show two plots that generalize (qualitatively) the importance of developing a wave simulation method (PE or otherwise) to apply to the in-ice radio propagation problem. In Figure~\ref{fig:spectrumDepth} we plot the received spectra as a function of depth for receivers in the top 200\,m of the ice, at a horizontal displacement of 1\,km from a transmitter buried 1050\,m in the ice. This is the spectrum of the first 100\,ns of a received pulse (with the same parameters as the pulses in the rest of this article). We restrict the spectrum to this window to reduce the effect of interference from direct + reflected signals (which is still evident in the near-surface pulses for which direct and reflected both lie within this 100\,ns window. Note that, in general, the direct pulse is affected less than the reflected pulse, spectrally). This is for an index of refraction profile corresponding to the first of two SPICE core profiles, applied at all $x$. While this may overestimate certain waveguide behavior, it is also an appropriate guess for a possible realistic ice profile, as discussed previously. Clearly visible near the surface are interference patterns (from direct and reflected pulses in close proximity) but also oscillations in the overall amplitude of the spectrum with depth. These oscillations gradually diminish with depth, with the spectrum becoming smoother as receivers get deeper, and are not an interference effect, but a result of propagation through the firn. There is a band at around $-140$\,m that shows some deviation from the smooth spectrum; this is perhaps some waveguide-like phenomena or interference from multiple scattering.

\begin{figure}[h]
   \centering
     \includegraphics[width=0.44\textwidth]{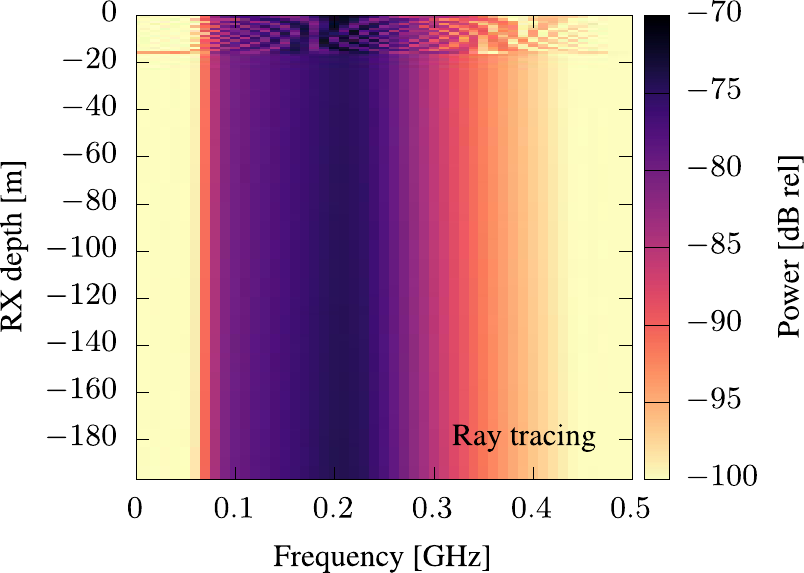}
     \caption{Spectrum of the first 100ns of a received signal for varying receiver (RX) depth using ray tracing. This is for a transmitter buried 1050\,m in the ice, displaced 1\,km in range from the receivers. The scale is dB (power) relative to the transmitter output.}
     \label{fig:spectrumDepthRT}

 \end{figure}

 Figure~\ref{fig:spectrumDepthRT} shows the same, but for RT methods. The interference near the surface is evident, but the oscillatory behavior as a function of depth is not.
 
\section{Implications for current and future experiments}
The initial results from this first work on in-ice PE suggest, in agreement with previous studies using FDTD~\cite{horizontal_propagation_fdtd}, that the effect of the ice on propagation may be significant, particularly for receivers in the firn. Specifically Figure~\ref{fig:spectrumDepth}, which shows a marked difference in the spectral content received at receivers shallow and deep from the same deep pulse, warrants further investigation into PE methods. 

Preliminarily, these results suggest that to avoid the effects of the firn, receivers should be placed below it, in the denser, more uniform ice from centuries past. These results further suggest, in agreement with previous studies, that ray-tracing alone over-simplifies propagation through the ice in ways that can have an effect on energy and position reconstruction for in-ice neutrino detectors, particularly for shallow receivers near the top of the domain in Figure~\ref{fig:spectrumDepth}. Receivers in this location reside in regions with the most strongly varying density profile (Figure~\ref{fig:nzprofile}). In this section we briefly discuss the potential implications for the two classes of radio based in-ice neutrino detectors, Askaryan and radar echo.

\subsection{Askaryan detectors}

Askaryan detectors rely on detecting the impulsive signals produced via the Askaryan effect. These impulsive signals are broadband, and achieve their maximum bandwidth directly at the preferred Cherenkov angle, which in deep ice is approximately 55$^{\circ}$ from the cascade axis. The spectral content of the received signal (via useful variables such as spectral slope~\cite{harm}) can elucidate where the received radio originated from with respect to this Cherenkov angle (radio at the Cherenkov angle is typically called ``on-cone'' because this radiation is radially symmetric on a cone at this angle about the cascade axis). A knowledge of where on the cone radio originated from allows for a signal's radio direction to be translated into a neutrino arrival direction, which can then be used to look for sources on the sky.

 \begin{figure}[h]
   \centering
     \includegraphics[width=0.45\textwidth]{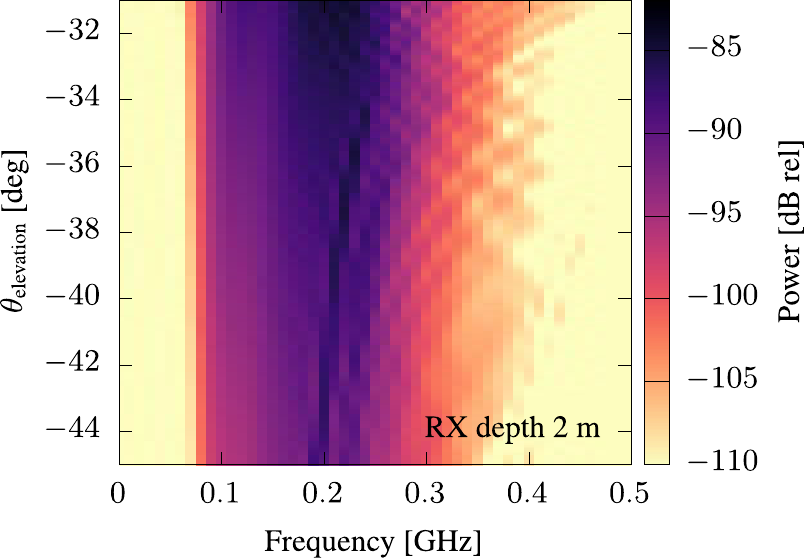}
     \caption{Spectrum of the first 100ns of a received signal (using PE) for a receiver at $(x,z)$=1350,-2\,m as a function of apparent arrival angle, measured with respect to the local horizontal of the receiver.}
     \label{fig:spectrumTheta2}

 \end{figure}

 \begin{figure}[h]
   \centering
     \includegraphics[width=0.45\textwidth]{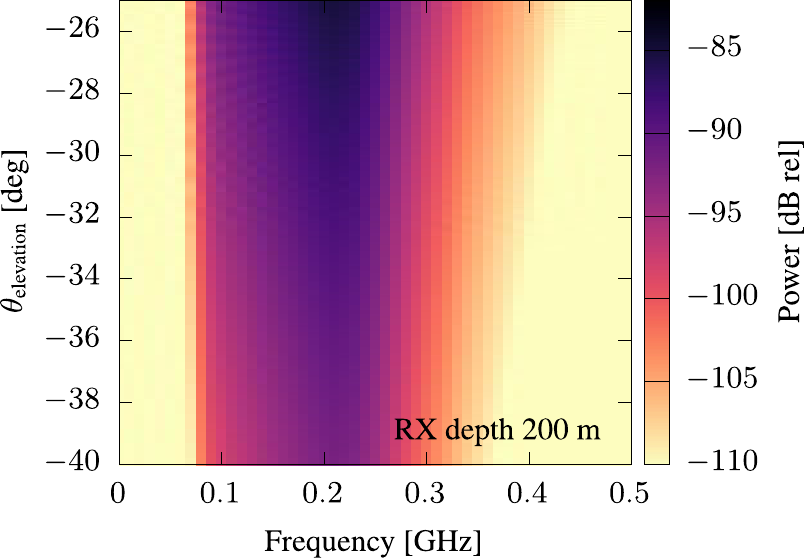}
     \caption{Spectrum of the first 100ns of a received signal (using PE) for a receiver at $(x,z)$=1350,-200\,m as a function of apparent arrival angle, measured with respect to the local horizontal of the receiver.}
     \label{fig:spectrumTheta200}

 \end{figure}

The received signal is affected by propagation through the ice in ways that affect the spectrum. For some geometries (such as Figure~\ref{td_compare_1350_2}) we see that the spectra are very different for signals simulated with PE and RT. If an idealized ice model is assumed, as in RT methods, then the received spectrum is assumed to be the source spectrum (i.e. no spectral changes occur as a result of propagation). As such, measurements of the spectral properties will lead to error in reconstructing the neutrino arrival direction, due to spectral changes that happened to the signal en-route. By studying and classifying these changes, reconstruction of neutrino arrival direction may be improved. 

 These results are not in tension with other studies, such as Ref.~\cite{arianna2020}, that show good vertex resolution for near-surface systems. These studies rely on the time of arrival difference between antennas (measured by cross-correlating the signals in different receivers) to point back to the source (in the case of Ref.~\cite{arianna2020} a transmitter lowered down the SPICE borehole), which does not rely on the shape of the pulse itself. The error in a measurement like this is then primarily down to the ability to identify the arrival time of a pulse. Signals such as those in Figure~\ref{td_compare_1350_2} may have different arrival times depending upon how this is measured, that being first rise above some threshold, peak amplitude of some envelope, or cross-correlation between channels, but this error results in a fairly small error in actual pointing (and in fact, if relying on the average of many pulses, tremendously precise measurements may be achieved in this way~\cite{ariannaRecoSnow}). However, the pulses in Figure~\ref{td_compare_1350_2} themselves, and more importantly their spectra, are significantly dissimilar such that if one wanted to extend a study from vertexing a neutrino interaction to an arrival direction study, which requires analysis of the spectrum to know where the event lies on the Cherenkov cone, these different pulses would provide different results. Therefore, while RT may allow for event vertexing with small error, the error on the actual neutrino arrival direction could be significant. A quantitative analysis of these effects is beyond the scope of this article, but we suggest based on these results that such analysis be performed. A more complete understanding of the spectral content of received signals can lead to an improvement in reconstruction over what is currently available in simulation codes. 

 Furthermore, if we look instead at the way an arrival spectrum changes for a fixed receiver as the transmitter is swept in depth, we see another subtle issue with assuming an idealized ice model. In Figures~\ref{fig:spectrumTheta2} and~\ref{fig:spectrumTheta200} we see that the signal in a single antenna does not change significantly except for the interference pattern from the direct and reflected timing and simple 1/$r$ changes. But this information, coupled with that in Figure~\ref{fig:spectrumDepth} is cause for concern, because one may not know whether they lie in the peak or trough of a large-scale oscillation, or waveguide-like structure that may artificially enhance or diminish certain frequencies, or the overall amplitude. Therefore, for antennas in this part of the ice, in-situ surveys with antennas at various depths and positions may be able to give a handle on this potential reconstruction systematic.

 \subsection{Implications for radar echo detection}

 The Radar Echo Telescope (RET) is a new detector technology that seeks to detect in-ice neutrinos via active radar sounding. When an UHE neutrino interacts in the ice, it produces a cascade (the same cascade that emits radio via the Askaryan effect). As this cascade moves relativistically through the medium, it ionizes the medium, leaving behind a short-lived cloud of ionization that can reflect incident radio waves. RET plans to deploy transmitting antennas to illuminate a volume of ice, as well as receiving antennas to monitor that same volume of ice, in order to detect neutrino-induced ionization deposits in this illuminated region. The radar echo method has recently been validated in the laboratory~\cite{t576_run2}, and a prototype in-situ detector is under development.

   \begin{figure}[h]
     \centering
       \includegraphics[width=0.45\textwidth]{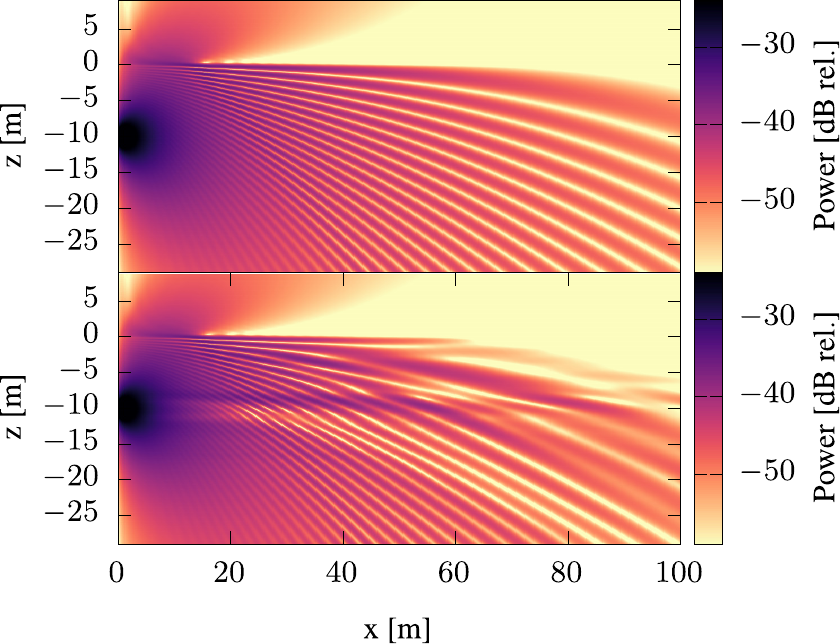}
     \caption{Comparison of peak power received for the functional (top) and data-driven (bottom) $n(x,z)$ profiles, for 350\,MHz continuous wave radio transmitted from a source 10\,m beneath the surface of the ice.}
     \label{fig:ret10m}

 \end{figure}

    \begin{figure}[h]
      \centering
        \includegraphics[width=0.45\textwidth]{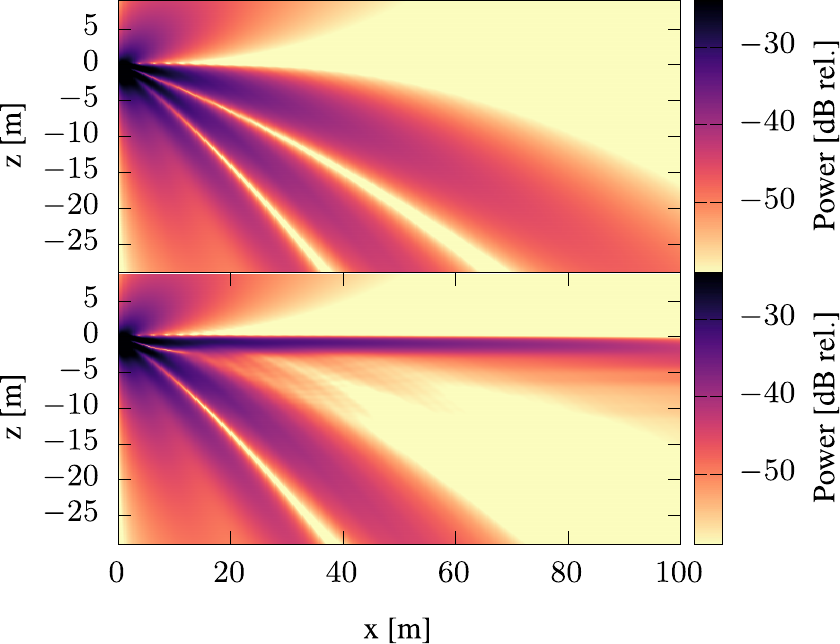}
     \caption{Comparison of peak power received for the functional (top) and data-driven (bottom) $n(x,z)$ profiles, for 350\,MHz continuous wave radio transmitted from a source 1\,m beneath the surface of the ice.}
     \label{fig:ret1m}

   \end{figure}

The received signal that RET will detect is very different from the Askaryan signal. Because RET uses active radar, the received signal is largely a function of the transmitted signal, coupled with the effects of the transmitter-cascade-receiver geometry. Similarly to Askaryan methods, however, the radar echo signal is subject to the effects of the ice density profile in transit; perhaps even more so because the radar echo signal makes 2 trips through the ice (transmitter---cascade, cascade---receiver). Therefore an understanding of the ice is essential to know what effects (geometry or the ice) are responsible for which signal characteristics, if the receivers are placed in such a way as to detect radio that will travel through the firn. Previous studies~\cite{radioscatter} have avoided this complication by stipulating that receivers and transmitter lie in deep, uniform ice.

 The NSF-supported pilot implementation of the RET is the Radar Echo Telescope for Cosmic Rays (RET-CR), which seeks to test the method in nature by detecting the {\it in-ice} cascade produced when an ultra-high-energy cosmic ray (UHECR) air shower impacts the ice. For a UHECR of sufficient energy ($\gtrsim10$\,PeV) and a high enough ice elevation ($\gtrsim$1.5\,km) a fraction $\gtrsim$10\% of the primary energy actually reaches the ice. This energy is tightly collimated around the cascade axis, and produces a dense, in-ice cascade near the surface of the ice. RET-CR will place a transmitter and receivers below the surface to test the radar echo method on this in-nature source. Because RET-CR will be situated near the surface, the effects of the $n(z)$ gradient will be pronounced. This present study, which qualitatively confirms (albeit with a different simulation method) the results from previous studies using FDTD~\cite{horizontal_propagation_fdtd}, highlights the importance of accurate modeling for near-surface propagation. For example, in Figure~\ref{fig:ret10m} we show two propagation maps for continuous-wave radio from a receiver 10\,m beneath the surface. We see that the interference pattern changes drastically depending upon which ice model is used. Therefore, accurate modeling is essential for RET to calculate an effective collection area, and to optimize the detector geometry. We also show, in Figure~\ref{fig:ret1m}, another important example of why wave methods like PE may be useful for such a study. In this Figure, we have placed a transmitter just 1\,m beneath the surface. In the lower panel we can see a very clear horizontally propagating mode, which is likely trapped in the deep $n(x,z)$ profile inversion visible in Figure~\ref{fig:nzprofile}, which is not present in the functional profile (top panel). While such effects may be local to only certain sites, it is nonetheless critical to have an understanding of the effects that different $n(x,z)$ profiles can have on propagation for an eventual radio system deployment.

\section{Discussion and Conclusions}

We note that this is the first application of PE methods to the in-ice problem, and the simulation is still in an introductory state. Considerations such as propagation at extremely large baselines, curvature of the earth's surface, reflections from the ice bottom, and ice surface roughness have not been thoroughly investigated nor included in the public version of the code. The effects of receiving antennas are not included; these results represent the field as it would arrive at the antenna. For a physical, extended receiver the effects could be pronounced, especially if the antenna is spread across several wavelengths in $z$, but this requires further study. The conclusions we present here are those that we feel confident to draw from the simulation as it currently exists, which show effects of propagation through the firn, vetted by FDTD, that are not modeled in current simulation codes.

We have presented the first application of parabolic equation methods to in-ice radio propagation, and have shown a modified split-step solution that reduces overall phase error for the in-ice case relative to existing solutions. We have shown that the received spectrum of a signal is affected by propagation through a medium with changing index of refraction, an effect which cannot be replicated with RT methods due to the infinite-frequency approximation. We showed that this effect is more pronounced when realistic $n(x,z)$ profiles are used, rather than the simplified functional $n(z)$ profile often used in ray tracing codes. The results herein are from a single RT implementation, courtesy of the ARA experiment simulation package (AraSim) and have been validated by a second RT implementation. We validated the PE method against FDTD simulations, and then made comparisons between PE and RT in realistically large domains.

In conclusion, we suggest that PE methods warrant further investigation, and that wave simulations are critical for simulating in-ice radio propagation. We see qualitatively that signals propagating through more of the firn have more significant distortion than for receivers below the firn in more uniform ice (see e.g. Figure~\ref{fig:spectrumDepth}). In agreement with previous studies, we find that the ice has a significant effect on spectral content and pulse shape. Though significant phase error is present in the PE methods presented here (seen in their deviation from FDTD), we can conclude that the PE formalism is useful for the in-ice problem. To that end, a more rigorous split-step approximation---or other PE solving routine entirely---would be welcome and useful. 
An understanding of these apparent ice effects on the received signal is critical for energy and direction reconstruction of neutrino events for in-ice radio neutrino detectors.

\section*{Acknowledgments}
This work was supported in part by the National Science Foundation under award numbers NSF/PHY-2012980 and NSF/PHY-2012989. We thank B.~Clark, C.~Glaser, R.~Lahmann, A.~Nelles, and R.~Prechelt for helpful comments. C.~Sbrocco was supported in part by an OSU Physics Undergraduate Summer Research Award. This work was also supported in part by the Flemish Foundation for Scientific Research FWO-12ZD920N, the European Research Council under the EU-ropean Unions Horizon 2020 research and innovation programme (grant agreement No 805486), and the Belgian Funds for Scientific Research (FRS-FNRS). DZB is grateful for support from the U.S. National Science Foundation-EPSCoR (RII Track-2 FEC, award ID 2019597). A. Connolly acknowledges support from NSF Award \#1806923. S.~Wissel was supported by NSF CAREER Awards \#1752922 and \#2033500. We thank the ARA Collaboration for making available the AraSim simulation program used in this work. Computing resources were provided by the Ohio Supercomputer Center. 
 
 \bibliography{paraPropArxivFinal}

\bigskip
 
 \appendix

 %\documentclass{article}
%\usepackage{amsmath}
%\begin{document}

\section{In-Ice Parabolic Equation}\label{pe}

\subsection{Derivation}
%We make a conventional scalar field approximation for systems with cylindrical symmetry. We factor the electric field into a polarization along the spherical polar direction $\hat{\theta}$, multiplied by a scalar field $\phi(\rho, z)$. The literature uses symbol $x$ for the cylindrical radius $\rho$, which we adopt to assist comparison. Thus $x$ is {\it not} the Cartesian-$x$ direction. 
We begin by assuming that we have a system with cylindrical symetry $(\rho, \theta, z)$, which is appropriate for many RF applications. The PE literature uses symbol $x$ for the cylindrical radius $\rho$, which we adopt to assist comparison, but note that this $x$ is {\it not} the Cartesian-$x$. For an arbitrary field $\psi$ which is polarized along $\theta$, the scalar wave equation for a field  (assuming an $e^{i\omega_0t}$ time dependence) is

\begin{equation}
  \label{eq:scalar1}
  (\nabla^2 - \frac{1}{v^2}\partial_t^2)\psi=0
\end{equation}
\begin{equation}
  \label{eq:scalar2}
  \nabla^2\psi + k_0^2n^2\psi=0
\end{equation}
where $k_0=\omega_0/c$ and $v=c/n$ for the vacuum speed of light $c$ and index of refraction $n$. Using

\begin{equation}
  \label{eq:reduced}
  \psi=\frac{1}{\sqrt{x}}u e^{ik_0x},
\end{equation}
we re-write Eq.~\ref{eq:scalar2} as

\begin{equation}
  \label{eq:u3d}
\partial_x^2 u + 2ik_0\partial_x u + \partial_z^2 u + k_0(n^2-1) u = -\frac{u}{4x^2}.
\end{equation}

The ansatz $u$ is called the ``reduced function'', and is used primarily because it allows for the convienent form of Eq.~\ref{eq:u3d} to be solved. By inspection, it can be seen that in the far field, that is, for large $x$, the r.h.s of Eq.~\ref{eq:u3d} approaches zero. We thus take this far-field approximation in what follows, which is valid in general for the problems of interest in radio propagation applications.

The parabolic approximation begins by formally factoring Eq.~\ref{eq:u3d},

\begin{equation}
  \label{eq:factored}
 \bigg[ \big(\partial_x + ik(1-Q)\big)\big(\partial_x + ik(1+Q)\big)\bigg]u=0,
\end{equation}
where we have taken the far-field $(x>>0)$ approximation, and introduced the pseudo-differential operator

\begin{equation}
Q=\sqrt{\frac{\partial_z^2}{k_0^2} + n^2}.
\end{equation}

We focus on solutions of the form

\begin{equation}
  u(x,z)=u_+(x,z)+u_-(x,z)
\end{equation}
\begin{equation}
  \partial_xu_{\pm}=-ik_0(1-Q)u_{\pm}
\end{equation}
where the $\pm$ indicate forward and backward propagating fields. The formal solution of the forward propagating field is% Individually, the $u_{\pm}$ have simple solutions. For applications where backward-propagating fields are negligible, we can throw away this part of the expression and focus our attention only on the forward propagating field. Focusing only on the forward-propagating field is called the paraxial approximation, and is the crux of the parabolic method. Then, for $u\sim u_+$, the solution has the form $\partial_x u=Au \rightarrow u=e^{Ax}$. 

\begin{equation}
  \label{eq:u}
u(x)=e^{ik_0x(-1+Q)}.
\end{equation}
Then, if we want to solve for $u(x+\Delta x)$, 

\begin{equation}
  \label{eq:stepping}
u(x+\Delta x)=e^{ik_0(x+\Delta x)(-1+Q)}=e^{ik_0\Delta x(-1+Q)}u(x),
\end{equation}
meaning that the solution at $u(x+\Delta x)$ only relies upon the previous solution $u(x)$. This means that the solution can be marched along in code, drastically simplifying the computational time needed to calculate a full domain. Focusing solely on the forward propagating field is the crux of several so-called parabolic approximations.

The operator $Q$ is non-local, and does not have simple analytic properties, so much care is needed when attempting to find ways to use it in numerical calculations. The simplest thing to do is to take the lowest order expansion of $Q$, which results in the expression

\begin{equation}
  \label{eq:spe}
\partial_z^2 u + 2ik_0\partial_x u + k_0^2(n^2-1)u=0,
\end{equation}
which is called the ``standard parabolic equation'' (SPE) in the literature. This approximation is quite good for small values of $n$ that change very slowly with height $z$, and for propagation angles within a few degrees of the propagation direction. However, for many problems (such as in-water acoustic propagation) it is useful to simulate propagation at wider angles and for a more strongly varying $n$. To that end, a different form of $Q$ attributed to Feit and Fleck~\cite{fleck1976time,feit1978light} is

\begin{equation}
  \label{eq:qff}
  Q_{FF}\sim\sqrt{1+\frac{\partial_z^2}{k_0^2}} + n-1.
\end{equation}
This expression has the same lowest-order expansion as the SPE, but differs at higher order. Notice this expression splits the operator into two parts, which may be classified as ``diffractive'', involving the partial derivative with respect to $z$, and ``refractive'', dealing with the index of refraction. Such splitting makes for efficient numerical solution. This splitting will be discussed below, when addressing the implementation of a solution in software.

Further inspection finds that this expression will be most accurate for $n\sim 1$, which may be suitable for the in-air radio or in-water acoustic cases, but this approximation is not suitable for the in-ice application. We therefore introduce a reference index value $n_0$ and an in-ice approximation for $Q$ that leads to efficient splitting,

\begin{equation}
  \label{eq:qice}
  Q_{ice}\sim\sqrt{1+\frac{\partial_z^2}{k_0^2}} + n\sqrt{1+\frac{1}{n_0^2}} - \sqrt{1 + \frac{n^2}{n_0^2}}.
\end{equation}
This also has the same lowest-order expansion as the SPE, differing at higher orders. We reach this approximation by using a reference wavenumber equal to that at the depth of the source. Meaning, instead of using $k_0=\omega/c$, we use $k_0=\omega n(z_0)/c$. Implementation shows improved agreement with FDTD calculations, as presented in this article. 

\subsection{Numerical Solution}

There are several methods to solve the parabolic equation. The form of the PE above, using $Q$ in the various approximate forms shown here, is solved using what are called ``split-step'' numerical methods in which the diffractive part of the field is solved for in Fourier space, and the refractive part of the field is solved for via simple multiplication. These solutions are computationally efficient, but the manipulation of $Q$ introduces error. The analysis of this error and desire to minimize it under certain use cases leads to the various approximations of $Q$.

The split-step method hinges on the Fourier identities $\mathcal{F}(\partial_z^2 u)=-k_z^2 \mathcal{F}(u)$ and $\mathcal{F}(\partial_x u)=\partial_x \mathcal{F}(u)$ in order to come to a solution that can be solved computationally, where $\mathcal{F}$ is the forward Fourier transform and $k_z$ is the vertical wavenumber bounded by $\pm\pi/\Delta z$. We begin by showing the solution for the SPE. First, we take the Fourier transform of Eq.~\ref{eq:spe}, which gives

\begin{equation}
  \label{eq:fudx}
  \partial_xU=\frac{ik_0}{2}\left[(n^2-1) -\frac{k_z^2}{k_0^2}\right]U,
\end{equation}
where $U(x, k_z) = \mathcal{F}\left(u(x,z)\right)$. By analogy to Eq.~\ref{eq:u}, we find

\begin{equation}
  \label{eq:fu}
  U(x, k_z)=\mathrm{exp}\left[\frac{ik_0}{2}\left((n^2-1) -\frac{k_z^2}{k_0^2}\right)x\right],
\end{equation}
and then using the same stepping idea as Eq.~\ref{eq:stepping}, we write
\begin{multline}
  \label{eq:fustep}
  U(x+\Delta x, k_z)=\\  \mathrm{exp}\left[\frac{ik_0\Delta x}{2}\left((n^2-1) -\frac{k_z^2}{k_0^2}\right)\right]U(x, k_z)
\end{multline}
From here, we take the inverse Fourier transform of Eq.~\ref{eq:fustep}, taking the part that does not rely on the transform variables outside of the transform, to arrive at

\begin{multline}
  \label{eq:spesplit}
  u(x+\Delta x, z)=\mathrm{exp}\left[\frac{ik_0\Delta x}{2}(n^2-1)\right]\\
  \mathcal{F}^{-1}\left\{\mathrm{exp}\left[\frac{-i\Delta x}{2}\frac{k_z^2}{k_0}\right]U(x, k_z)\right\}.
\end{multline}
This is the split-step solution to the SPE, Eq.~\ref{eq:spe}. By a similar procedure, we use $Q_{ice}$ from Eq.~\ref{eq:qice} to arrive at the form of the split-step solution used in this article and in paraPropPython,

\begin{multline}
  \label{eq:parabolic_final_app}
  u(x+\Delta x, z)=\textrm{exp}\bigg[ik_0\left(n\sqrt{1+ \frac{1}{n_0^2}}-\sqrt{1+\frac{n^2}{n_0^2}}\right)\Delta x\bigg]\\
  \mathcal{F}^{-1}\bigg\{\textrm{exp}\bigg[-ik_0\Delta x\sqrt{1-\frac{k_z^2}{k_0^2}}+1\bigg]U(x, k_z)\bigg\},     
\end{multline}
where $n=n(x, z)$ is written for clarity.

\subsection{Implementation notes}
Following~\cite{levy,apaydin2010split, apaydin2017radio}, paraPropPython uses artificial filtering at the top and bottom of the simulation domain to eliminate artificial reflections. The simulation domain in paraPropPython is twice the maximum user-specified depth (e.g. it simulates a region above the ice the same height in $z$ as the simulation domain is deep in $z$) plus a small buffer region above and below. Future releases will allow the user to specify the exact simulation domain, to speed up computation, if the in-air portion is not needed, but in all cases this buffer region above and below is necessary. For this reason, reflections from the bottom of the ice are not considered currently in the code.

Surface roughness is also not included at this current time, though we expect this to have some impact on the reflected signal properties. Surface roughness will vary from site-to-site, and is an important feature to include in future versions of the code.

The default implementation of paraPropPython is to use the split-step approximation outlined here. The user can also specify other split-step approximations for comparison purposes, including the standard wide-angle, and the original Feit and Fleck splitting.

\section{Source Modeling}\label{source}

Many parabolic equations model a source as a Gaussian beam of some width, to simulate a directional antenna over the surface of the earth. We are interested in what the propagation might look like from a dipole source, since it is much simpler to bury a dipole deep into the ice than it is a high-gain antenna. We therefore define a dipole at a depth $z_0$ with halves of approximately $L=\lambda/4$. We approximate this distance as closely as possible to the nearest grid spacing in the simulation. We then define the reduced field within this source region, for a vertically polarized dipole, by

\begin{equation}
  u(0, z_0-L:z_0+L)=A [\hat{n} \times \hat{\epsilon} \times \hat{n}]_z ,
\end{equation}

where $A$ is a (complex) amplitude, $\hat{n}$ is a unit vector that points out radially from the dipole (such that $\hat{n}_x(z_0-L)=0$, $\hat{n}_x(z_0)=1$, and $\hat{n}_x(z_0+L)=0$), $\epsilon$ is the polarization vector of the antenna ($\epsilon=(0\hat{x}, 0\hat{y}, 1\hat{z})$ for a vertically polarized antenna), and the subscript $z$ indicates that the reduced field $u$ corresponds to the $\hat{z}$ component of this expression. Such a formalism results in the typical sin$^2\theta$ pattern of a dipole antenna. In 2-d, this results in a cross-sectional slice of the antenna beam pattern along the halves of the source. Such a source shows better agreement with a dipole FDTD source than sources typically used in the literature.

%\end{document}

\end{document}